\documentclass[preprint,12pt]{elsarticle}

\usepackage{amsmath}
\usepackage{amsfonts}
\usepackage{epsfig}
\usepackage{graphics,color}
\usepackage{verbatim}
\usepackage{tabularx}
\usepackage{theorem}
\usepackage{multirow}
\usepackage{xspace}
\usepackage{algorithmic}
\usepackage[linesnumbered,ruled,vlined]{algorithm2e}
\usepackage[latin1]{inputenc}
\usepackage{subfigure}
\usepackage{ctable}
\usepackage{balance}
\usepackage{acronym}
\usepackage{graphicx}
\usepackage{color}

\graphicspath{{./}}

\usepackage{amssymb}
\usepackage{url}

\acrodef{SLA}{Service Level Agreement}
\acrodef{QoS}{Quality of Service}
\acrodef{AaaS}{Analytics as a Service}
\acrodef{BDaaS}{Big Data as a Service} 
\acrodef{IaaS}{Infrastructure as a Service}
\acrodef{PaaS}{Platform as a Service}
\acrodef{SaaS}{Software as a Service}
\acrodef{MaaS}{Model as a Service}
\acrodef{SME}{Small and Medium Enterprise}
\acrodef{CaaaS}{Continuous analytics as a Service}
\acrodef{S3}{Simple Storage Service}
\acrodef{HDFS}{Hadoop Distributed File System}
\acrodef{GPFS}{General Parallel File System}
\acrodef{PVFS}{Parallel Virtual File System}
\acrodef{EDW}{Enterprise Data Warehouse}
\acrodef{BI}{Business Intelligence}
\acrodef{CRM}{Customer Relationship Management}
\acrodef{EMR}{Elastic MapReduce}
\acrodef{EC2}{Elastic Compute Cloud}
\acrodef{GAE}{Google App Engine}
\acrodef{OLAP}{Online Analytical Processing}
\acrodef{DaaS}{Data as a Service}
\acrodef{PMML}{Predictive Model Markup Language}
\acrodef{RDBMS}{Relational Database Management System}
\acrodef{MDC}{Mobile Data Challenge}
\acrodef{GFS}{Google File System}
\acrodef{MML}{Medical Markup Language}
\acrodef{DCG}{Distributed Computation Graphs}
\acrodef{MMOG}{Massively Multiplayer Online Game}
\acrodef{BPEL}{Business Process Execution Language}
\acrodef{DAG}{Direct Acyclic Graph}

\journal{Journal of Parallel and Distributed Computing}

\begin{document}

\begin{frontmatter}



\title{Big Data Computing and Clouds:\\ Trends and Future Directions}

\author[inria]{Marcos D. Assun\c{c}\~{a}o\corref{cor1}}
\author[unimelb]{Rodrigo N. Calheiros}
\author[ibm]{Silvia Bianchi}
\author[ibm]{Marco~A.~S.~Netto}
\author[unimelb]{Rajkumar Buyya\corref{cor1}}

\address[inria]{INRIA, LIP, ENS de Lyon, France}
\address[unimelb]{The University of Melbourne, Australia}
\address[ibm]{IBM Research, Brazil}
\cortext[cor1]{Corresponding authors: assuncao@acm.org, rbuyya@unimelb.edu.au}

\begin{abstract}
This paper discusses approaches and environments for carrying out analytics on Clouds for Big Data applications. It revolves around four important areas of analytics and Big Data, namely (i) data management and supporting architectures; (ii) model development and scoring; (iii) visualisation and user interaction; and (iv) business models. Through a detailed survey, we identify possible gaps in technology and provide recommendations for the research community on future directions on Cloud-supported Big Data computing and analytics solutions.
\end{abstract}


\end{frontmatter}


\section{Introduction}

Society is becoming increasingly more instrumented and as a result, organisations are producing and storing vast amounts of data. Managing and gaining insights from the produced data is a challenge and key to competitive advantage. Analytics solutions that mine structured and unstructured data are important as they can help organisations gain insights not only from their privately acquired data, but also from large amounts of data publicly available on the Web \cite{SchommDataMarket:2013}. The ability to cross-relate private information on consumer preferences and products with information from tweets, blogs, product evaluations, and data from social networks opens a wide range of possibilities for organisations to understand the needs of their customers, predict their wants and demands, and optimise the use of resources. This paradigm is being popularly termed as Big Data.

Despite the popularity on analytics and Big Data, putting them into practice is still a complex and time consuming endeavour. As Yu~\cite{YuBigData} points out, Big Data offers substantial value to organisations willing to adopt it, but at the same time poses a considerable number of challenges for the realisation of such added value. An organisation willing to use analytics technology frequently acquires expensive software licenses; employs large computing infrastructure; and pays for consulting hours of analysts who work with the organisation to better understand its business, organise its data, and integrate it for analytics \cite{SunBusinessModels:2011}. This joint effort of organisation and analysts often aims to help the organisation understand its customers' needs, behaviours, and future demands for new products or marketing strategies. Such effort, however, is generally costly and often lacks flexibility. Nevertheless, research and application of Big Data are being extensively explored by governments, as evidenced by initiatives from USA~\cite{USABigData} and UK~\cite{UKBigData}; by academics, such as the bigdata@csail initiative from MIT~\cite{MIT}; and by companies such as Intel~\cite{intel}.

Cloud computing has been revolutionising the IT industry by adding flexibility to the way IT is consumed, enabling organisations to pay only for the resources and services they use. In an effort to reduce IT capital and operational expenditures, organisations of all sizes are using Clouds to provide the resources required to run their applications. Clouds vary significantly in their specific technologies and implementation, but often provide infrastructure, platform, and software resources as services~\cite{buyya2009cloud,ArmbrustCloud:2009}.

The most often claimed benefits of Clouds include offering resources in a pay-as-you-go fashion, improved availability and elasticity, and cost reduction. Clouds can prevent organisations from spending money maintaining peak-provisioned IT infrastructure they are unlikely to use most of the time. Whilst at first glance the value proposition of Clouds as a platform to carry out analytics is strong, there are many challenges that need to be overcome to make Clouds an ideal platform for scalable analytics.

In this article we survey approaches, environments, and technologies on areas that are key to Big Data analytics capabilities and discuss how they help building analytics solutions for Clouds. We focus on the most important technical issues on enabling Cloud analytics, but also highlight some of the non-technical challenges faced by organisations that want to provide analytics as a service in the Cloud. In addition, we describe a set of gaps and recommendations for the research community on future directions on Cloud-supported Big Data computing.

\section{Background and Methodology}
\label{sec:background}

Organisations are increasingly generating large volumes of data as result of instrumented business processes, monitoring of user activity \cite{attentionshoppers:2013,CiscoMSE:2012}, web site tracking, sensors, finance, accounting, among other reasons. With the advent of social network Web sites, users create records of their lives by daily posting details of activities they perform, events they attend, places they visit, pictures they take, and things they enjoy and want. This data deluge is often referred to as Big Data \cite{McAfeeBigData:2012,FranksBigData:2012,BellBigData:2009}; a term that conveys the challenges it poses on existing infrastructure in respect to storage, management, interoperability, governance, and analysis of the data.

In today's competitive market, being able to explore data to understand customer behaviour, segment customer base, offer customised services, and gain insights from data provided by multiple sources is key to competitive advantage. Although decision makers would like to base their decisions and actions on insights gained from this data \cite{DavenportAnalytics:2010}, making sense of data, extracting non obvious patterns, and using these patterns to predict future behaviour are not new topics. Knowledge Discovery in Data (KDD) \cite{FayyadKDD:1996} aims to extract non obvious information using careful and detailed analysis and interpretation. Data mining \cite{WittenDataMining:2011,KingBuyDataMining:2008}, more specifically, aims to discover previously unknown interrelations among apparently unrelated attributes of datasets by applying methods from several areas including machine learning, database systems, and statistics. Analytics comprises techniques of KDD, data mining, text mining, statistical and quantitative analysis, explanatory and predictive models, and advanced and interactive visualisation to drive decisions and actions \cite{DavenportAnalytics:2010,davenport2007competing,grossman2009what}.


\begin{figure}[!ht]
\center
\includegraphics[width=0.95\textwidth]{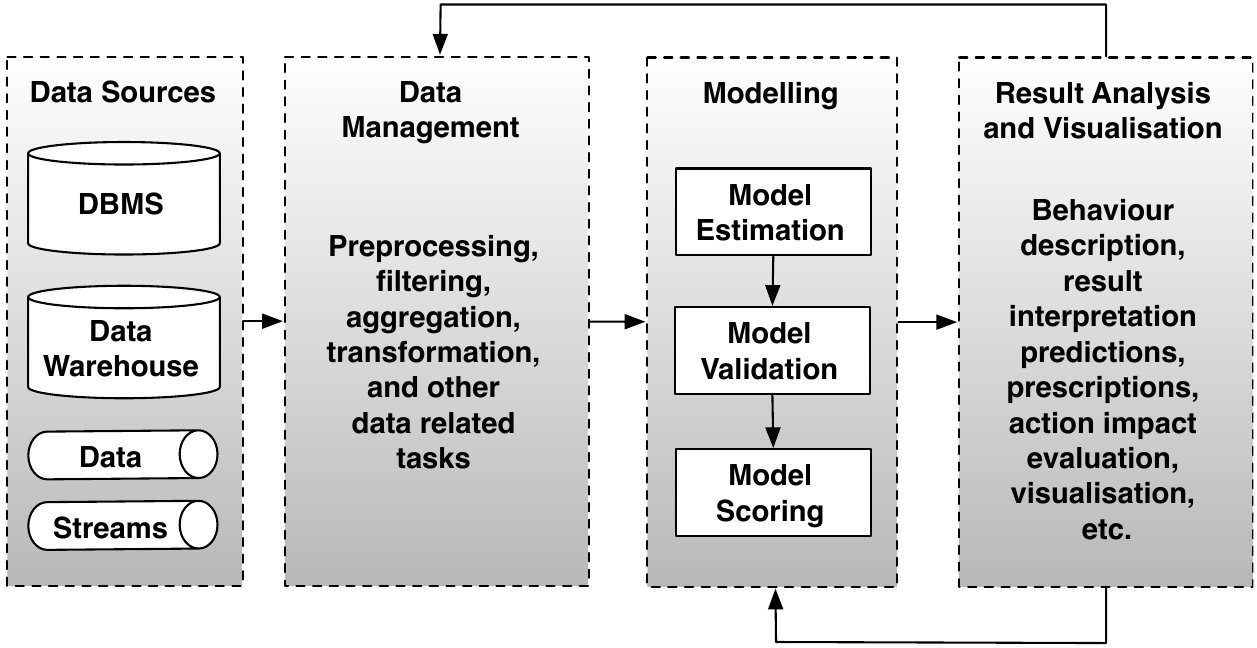}
\caption{Overview of the analytics workflow for Big Data.}
\label{fig:analytics_workflow}
\end{figure}

Figure~\ref{fig:analytics_workflow} depicts the common phases of a traditional analytics workflow for Big Data. Data from various sources, including databases, streams, marts, and data warehouses, are used to build models. The large volume and different types of the data can demand pre-processing tasks for integrating the data, cleaning it, and filtering it. The prepared data is used to train a model and to estimate its parameters. Once the model is estimated, it should be validated before its consumption. Normally this phase requires the use of the original input data and specific methods to validate the created model. Finally, the model is consumed and applied to data as it arrives. This phase, called model scoring, is used to generate predictions, prescriptions, and recommendations. The results are interpreted and evaluated, used to generate new models or calibrate existing ones, or are integrated to pre-processed data. 

\begin{figure}[!ht]
\center
\includegraphics[width=0.85\textwidth]{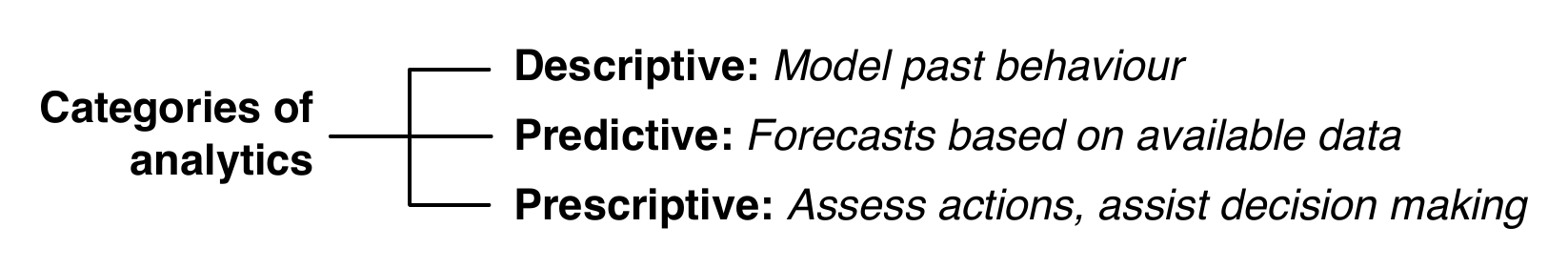}
\caption{Categories of analytics.}
\label{fig:analytics_categories}
\end{figure}

Analytics solutions can be classified as descriptive, predictive, or prescriptive as illustrated in Figure \ref{fig:analytics_categories}. Descriptive analytics uses historical data to identify patterns and create management reports; it is concerned with modelling past behaviour. Predictive analytics attempts to predict the future by analysing current and historical data. Prescriptive solutions assist analysts in decisions by determining actions and assessing their impact regarding business objectives, requirements, and constraints.

Despite the hype about it, using analytics is still a labour intensive endeavour. This is because current solutions for analytics are often based on proprietary appliances or software systems built for general purposes. Thus, significant effort is needed to tailor such solutions to the specific needs of the organisation, which includes integrating different data sources and deploying the software on the company's hardware (or, in the case of appliances, integrating the appliance hardware with the rest of the company's systems) \cite{SunBusinessModels:2011}. Such solutions are usually developed and hosted on the customer's premises, are generally complex, and their operations can take hours to execute. Cloud computing provides an interesting model for analytics, where solutions can be hosted on the Cloud and consumed by customers in a pay-as-you-go fashion. For this delivery model to become reality, however, several technical issues must be addressed, such as data management, tuning of models, privacy, data quality, and data currency. 

This work highlights technical issues and surveys existing work on solutions to provide analytics capabilities for Big Data on the Cloud. Considering the traditional analytics workflow presented in Figure~\ref{fig:analytics_workflow}, we focus on key issues in the phases of an analytics solution. With Big Data it is evident that many of the challenges of Cloud analytics concern data management, integration, and processing. Previous work has focused on issues such as data formats, data representation, storage, access, privacy, and data quality. Section~\ref{sec:data_mgmt} presents existing work addressing these challenges on Cloud environments.  In Section \ref{sec:models}, we elaborate on existing models to provide and evaluate data models on the Cloud. Section~\ref{sec:visualisation} describes solutions for data visualisation and  customer interaction with analytics solutions provided by a Cloud. We also highlight some of the business challenges posed by this delivery model when we discuss service structures, service level agreements, and business models. Security is certainly a key challenge for hosting analytics solutions on public Clouds. We consider, however, that security is an extensive topic and would hence deserve a study of its own. Therefore, security and evaluation of data correctness \cite{WangCorrectness:2012} are out of scope of this survey. 

\section{Data Management}
\label{sec:data_mgmt}

One of the most time-consuming and labour-intensive tasks of analytics is preparation of data for analysis; a problem often exacerbated by Big Data as it stretches existing infrastructure to its limits. Performing analytics on large volumes of data requires efficient methods to store, filter, transform, and retrieve the data. Some of the challenges of deploying data management solutions on Cloud environments have been known for some time \cite{AbadiDataMgmt:2009,SakrSurvey:2011,KatzInfras:2012}, and solutions to perform analytics on the Cloud face similar challenges. Cloud analytics solutions need to consider the multiple Cloud deployment models adopted by enterprises, where Clouds can be for instance:

\begin{itemize}
\item \textit{Private}: deployed on a private network, managed by the organisation itself or by a third party. A private Cloud is suitable for businesses that require the highest level of control of security and data privacy. In such conditions, this type of Cloud infrastructure can be used to share the services and data more efficiently across the different departments of a large enterprise. 

\item \textit{Public}: deployed off-site over the Internet and available to the general public. Public Cloud offers high efficiency and shared resources with low cost. The analytics services and data management are handled by the provider and the quality of service (\textit{e.g.} privacy, security, and availability) is specified in a contract. Organisations can leverage these Clouds to carry out analytics with a reduced cost or share insights of public analytics results. 

\item \textit{Hybrid}: combines both Clouds where additional resources from a public Cloud can be provided as needed to a private Cloud. Customers can develop and deploy analytics applications using a private environment, thus reaping benefits from elasticity and higher degree of security than using only a public Cloud.
\end{itemize}

Considering the Cloud deployments, the following scenarios are generally envisioned regarding the availability of data and analytics models \cite{Krishnainfosys:2012}: (i) data and models are private; (ii) data is public, models are private; (iii) data and models are public; and (iv) data is private, models are public. Jensen \textit{et al.} \cite{IBMBusinessCloud:2012} advocate on deployment models for Cloud analytics solutions that vary from solutions using privately hosted software and infrastructure, to private analytics hosted on a third party infrastructure, to public model where the solutions are hosted on a public Cloud.

Different from traditional Cloud services, analytics deals with high-level capabilities that often demand very specialised resources such as data and domain experts' analysis skills. For this reason, we advocate that under certain business models---especially those where data and models reside on the provider's premises---not only ordinary Cloud services, but also the skills of data experts need to be managed. To achieve economies of scale and elasticity, Cloud-enabled Big Data analytics needs to explore means to allocate and utilise these specialised resources in a proper manner. The rest of this section discusses existing solutions on data management irrespective of where data experts are physically located, focusing on storage and retrieval of data for analytics; data diversity, velocity and integration; and resource scheduling for data processing tasks.

\subsection{Data Variety and Velocity}

\begin{figure}[!ht]
\center
\includegraphics[width=0.85\textwidth]{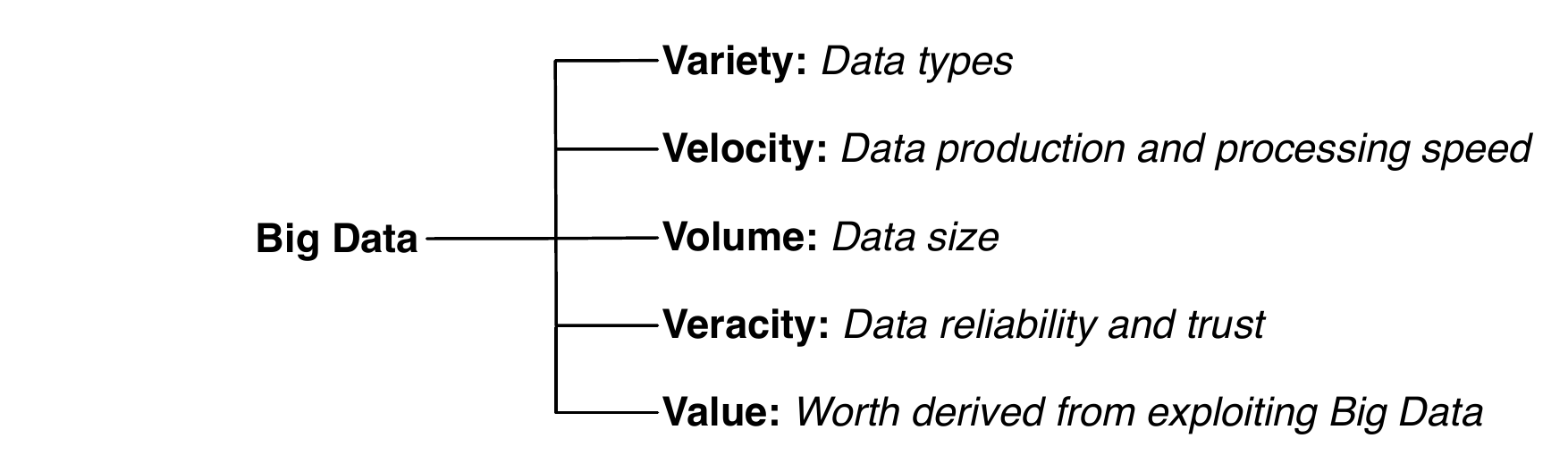}
\caption{Some `Vs' of Big Data.}
\label{fig:4vs}
\end{figure}

Big Data is characterised by what is often referred to as a multi-V model, as depicted in Figure~\ref{fig:4vs}. Variety represents the data types, velocity refers to the rate at which the data is produced and processed, and volume defines the amount of data. Veracity refers to how much the data can be trusted given the reliability of its source~\cite{YuBigData}, whereas value corresponds the monetary worth that a company can derive from employing Big Data computing. Although the choice of Vs used to explain Big Data is often arbitrary and varies across reports and articles on the Web---\textit{e.g.} as of writing Viability is becoming a new V---variety, velocity, and volume~\cite{RussomBigDataAnalytics:2011,ibm2012understanding} are the items most commonly mentioned.

Regarding Variety, it can be observed that over the years, substantial amount of data has been made publicly available for scientific and business uses. Examples include repositories with government statistics\footnote{http://www.data.gov}; historical weather information and forecasts; DNA sequencing; information on traffic conditions in large metropolitan areas; product reviews and comments; demographics \cite{PivotLink}; comments, pictures, and videos posted on social network Web sites; information gathered using citizen-science platforms \cite{BonneyCitizenScience2014}; and data collected by a multitude of sensors measuring various environmental conditions such as temperature, air humidity, air quality, and precipitation.

An example illustrating the need for such a variety within a single analytics application is the Eco-Intelligence~\cite{ZhangEcocity:2010} platform. Eco-Intelligence was designed to analyse large amounts of data to support city planning and promote more sustainable development. The platform aims to efficiently discover and process data from several sources, including sensors, news, Web sites, television and radio, and exploit information to help urban stakeholders cope with the highly dynamics of urban development.  In a related scenario, the \ac{MDC} was created aimed at generating innovations on smartphone-based research, and to enable community evaluation of mobile data analysis methodologies \cite{LaurilaNokiaMDC:2012}. Data from around 200 users of mobile phones was collected over a year as part of the Lausanne Data Collection Campaign. Another related area benefiting from analytics is \acp{MMOG}. CAMEO \cite{IosupCAMEO:2010} is an architecture for continuous analytics for MMOGs that uses Cloud resources for analysis of tasks. The architecture provides mechanisms for data collection and continuous analytics on several factors such as understanding the needs of the game community.

Data is also often available for sale in the format of research and technical reports, market segment and financial analyses, among other means. This data can be used by various applications, for instance, to improve the living conditions in large cities, to provide better quality services, to optimise the use of natural resources\footnote{Sense-T laboratory: http://www.sense-t.org.au/about/the-big-picture}, and to prevent or manage response to unplanned events.

\begin{figure}[!ht]
\center
\includegraphics[width=0.85\textwidth]{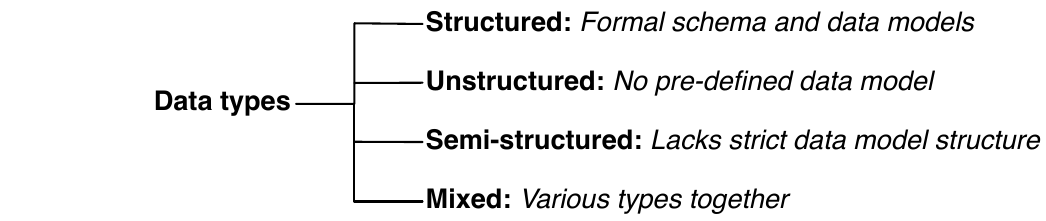}
\caption{Variety of data.}
\label{fig:data_variety}
\end{figure}

Handling and analysing this data poses several challenges as it can be of different types (Figure \ref{fig:data_variety}). It is argued that a large part of data produced today is either \textit{unstructured} or \textit{semi-structured}.

\begin{figure}[!ht]
\center
\includegraphics[width=0.85\textwidth]{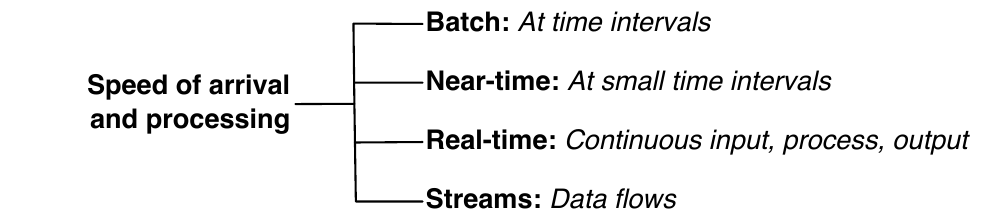}
\caption{Velocity of data.}
\label{fig:data_velocity}
\end{figure}

Considering data Velocity, it is noticed that, to complicate matters further, data can arrive and require processing at different speeds, as illustrated in Figure \ref{fig:data_velocity}. Whist for some applications, the arrival and processing of data can be performed in batch, other analytics applications require continuous and real-time analyses, sometimes requiring immediate action upon processing of incoming data streams. For instance, to provide active management for data centres, Wang \textit{et al.} \cite{WangFlexibleArch:2011} present an architecture that integrates monitoring and analytics. The proposed architecture relies on \ac{DCG} that are created to implement the desired analytics functions. The motivating use cases consist in scenarios where information can be collected from monitored equipments and services, and once a potential problem is identified, the system can instantiate \acp{DCG} to collect further information for analytics.

Increasingly often, data arriving via streams needs to be analysed and compared against historical information. Different data sources may use their own formats, which makes it difficult to \textit{integrate} data from multiple sources in an analytics solution. As highlighted in existing work \cite{fisher2012interactions}, \textit{standard formats and interfaces} are crucial so that solution providers can benefit from economies of scale derived from data integration capabilities that address the needs of a wide range of customers.

The rest of our discussion on data management for Cloud analytics surrounds three Vs of Big Data, namely Variety, Velocity and Volume. We survey solutions on how this diverse data is stored, how it can be integrated and how it is often processed. The discussion on visualisation also explores Velocity by highlighting factors such as interactivity and batch based visualisation. Although the other Vs of Big Data are important, we consider that some of them, as discussed earlier, deserve a study of their own, such as data Veracity. Other Vs are subjective; Value may depend on how efficient a company employs the analytics solutions at hand. Besides the V attributes, Big Data analytics also shares concerns with other data-related disciplines, and thus can directly benefit from the body of knowledge developed in the last years on such established subjects. This is the case of issues such as data quality~\cite{redman1997dataquality} and data provenance~\cite{Moreau2008provenance}.

\subsection{Data Storage}
\label{subsec:data_storage}

Several solutions were proposed to store and retrieve large amounts of data demanded by Big Data, some of which are currently used in Clouds. Internet-scale file systems such as the \ac{GFS} \cite{GhemawatGFS:2003} attempt to provide the robustness, scalability, and reliability that certain Internet services need. Other solutions provide object-store capabilities where files can be replicated across multiple geographical sites to improve redundancy, scalability, and data availability. Examples include Amazon \ac{S3}\footnote{http://aws.amazon.com/s3/}, Nirvanix Cloud Storage\footnote{http://www.nirvanix.com}, OpenStack Swift\footnote{http://swift.openstack.org} and Windows Azure Binary Large Object (Blob) storage\footnote{http://www.windowsazure.com/en-US/services/data-management/}. Although these solutions provide the scalability and redundancy that many Cloud applications require, they sometimes do not meet the concurrency and performance needs of certain analytics applications.

One key aspect in providing performance for Big Data analytics applications is the \emph{data locality}. This is because the volume of data involved in the analytics makes it prohibitive to transfer the data to process it. This was the preferred option in typical high performance computing systems: in such systems, that typically concern performing CPU-intensive calculations over a moderate to medium volume of data, it is feasible to transfer data to the computing units, because the ratio of data transfer to processing time is small. Nevertheless, in the context of Big Data, this approach of moving data to computation nodes would generate large ratio of data transfer time to processing time. Thus, a different approach is preferred, where computation is moved to where the data is. The same approach of exploring data locality was explored previously in scientific workflows~\cite{Deelman:2008} and in Data Grids~\cite{Venugopal2006datagrid} 

In the context of Big Data analytics, MapReduce presents an interesting model where data locality is explored to improve the performance of applications. Hadoop, an open source MapReduce implementation, allows for the creation of clusters that use the \ac{HDFS} to partition and replicate datasets to nodes where they are more likely to be consumed by mappers. In addition to exploiting concurrency of large numbers of nodes, \ac{HDFS} minimises the impact of failures by replicating datasets to a configurable number of nodes. It has been used by Thusoo~\textit{et al.}~\cite{thusoo2010data} to develop an analytics platform to process Facebook's large data sets. The platform uses Scribe to aggregate logs from  Web servers and then exports them to \ac{HDFS} files and uses a Hive-Hadoop cluster to execute analytics jobs. The platform includes replication and compression techniques and columnar compression of Hive\footnote{http://hive.apache.org} to store large amounts of data.

Among the drawbacks of Cloud storage techniques and MapReduce implementations, there is the fact that they require the customer to learn a new set of APIs to build analytics solutions for the Cloud. To minimise this hurdle, previous work has also investigated POSIX-like file systems for data analytics. As an example, Ananthanarayan \textit{et al.}~\cite{ananthanarayana2009cloud} adapted POSIX-based cluster file systems to be used as data storage for Cloud analytics applications. By using the concept of meta-blocks, they demonstrated that IBM's \ac{GPFS}  \cite{SchmuckGPFS:2002} can match the read performance of \ac{HDFS}. A meta-block is a consecutive set of data blocks that are allocated in the same disk, thus guaranteeing contiguity. The proposed approach explores the trade-off between different block sizes, where meta-blocks minimise seek overhead in MapReduce applications, whereas small blocks reduce pre-fetch overhead and improves cache management for ordinary applications. Tantisiriroj \textit{et al.} \cite{tantisiriroj2011duality} compared the \ac{PVFS}~\cite{CarnsPVFS:2000} against \ac{HDFS}, where they observed that \ac{PVFS} did not present significant improvement in completion time and throughput compared to \ac{HDFS}.

Although a large part of the data produced nowadays is unstructured, relational databases have been the choice most organisations have made to store data about their customers, sales, and products, among other things. As data managed by traditional DBMS ages, it is moved to data warehouses for analysis and for sporadic retrieval. Models such as MapReduce are generally not the most appropriate to analyse such relational data. Attempts have been made to provide hybrid solutions that incorporate MapReduce to perform some of the queries and data processing required by DBMS's \cite{AbadiDataMgmt:2009}. Cohen \textit{et al.}~\cite{cohen2009mad} provide a parallel database design for analytics that supports SQL and MapReduce scripting on top of a DBMS to integrate multiple data sources. A few providers of analytics and data mining solutions, by exploring models such as MapReduce, are migrating some of the processing tasks closer to where the data is stored, thus trying to minimise surpluses of siloed data preparation, storage, and processing \cite{KobielusInDatabase:2009}. Data processing and analytics capabilities are moving towards \acp{EDW}, or are being deployed in data hubs \cite{IBMBusinessCloud:2012} to facilitate reuse across various data sets.

In respect to \ac{EDW}, some Cloud providers offer solutions that promise to scale to one petabyte of data or more. Amazon Redshift \cite{AmazonRedshift}, for instance, offers columnar storage and data compression and aims to deliver high query performance by exploring a series of features, including a massively parallel processing architecture using high performance hardware, mesh networks, locally attached storage, and zone maps to reduce the I/O required by queries. Amazon Data Pipeline \cite{AmazonDatapipeline} allows a customer to move data across different Amazon Web Services, such as \ac{EMR} \cite{AmazonEMR} and DynamoDB \cite{AmazonDynamoDB}, and hence compose the required analytics capabilities.

Another distinctive trend in Cloud computing is the increasing use of NoSQL databases as the preferred method for storing and retrieving information. NoSQL adopts a non-relational model for data storage. Leavitt argues that non-relational models have been available for more than 50 years in forms such as object-oriented, hierarchical, and graph databases, but recently this paradigm started to attract more attention with models such as key-store, column-oriented, and document-based stores~\cite{Leavitt:2010}. The causes for such raise in interest, according to Levitt, are better performance, capacity of handling unstructured data, and suitability for distributed environments~\cite{Leavitt:2010}. 

Han \emph{et al.} \cite{Han2011NOSQLSurvey} presented a survey of NoSQL databases with emphasis on their advantages and limitations for Cloud computing. The survey classifies NoSQL systems according to their capacity in addressing different pairs of CAP (consistency, availability, partitioning). The survey also explores the data model that the studied NoSQL systems support.

Hecht and Jablonski \cite{HechtNoSQLEval} compared different NoSQL systems in regard to supported data models, types of query supported, and support for concurrency, consistency, replication, and partitioning. Hecht and Jablonski concluded that there are big differences among the features of different technologies, and there is no single system that would be the most suitable for every need. Therefore, it is important for adopters to understand the requirements of their applications and the capabilities of different systems so the system whose features better match their needs is selected~\cite{HechtNoSQLEval}.


\subsection{Data Integration Solutions}

Forrester Research published a technical report that discusses some of the problems that traditional \ac{BI} faces \cite{KobielusInDatabase:2009}, highlighting that there is often a surplus of siloed data preparation, storage, and processing. Authors of the report envision some data processing and Big Data analytics capabilities being migrated to the \ac{EDW}, hence freeing organisations from unnecessary data transfer and replication and the use of disparate data-processing and analysis solutions. Moreover, as discussed earlier, they advocated that analytics solutions will increasingly expose data processing and analysis features via MapReduce and SQL-MR-like interfaces. SAP HANA One~\cite{SapHanaOne}, as an example, is a in-memory platform hosted by Amazon Web Services that provides real-time analytics for SAP applications. HANA One also offers a SAP data integrator to load data from \ac{HDFS} and Hive-accessible databases.

\acp{EDW} or Cloud based data warehouses, however, create certain issues in respect to data integration and the addition of new data sources. Standard formats and interfaces can be essential to achieve economies of scale and meet the needs of a large number of customers \cite{fisher2012interactions}. Some solutions attempt to address some of these issues \cite{PivotLink,Birst}. Birst \cite{Birst} provides composite spaces and space inheritance, where a composite space integrates data from one or more parent spaces with additional data added to the composite space. Birst provides a \ac{SaaS} solution that offers analytics functionalities on a subscription model; and appliances with the business analytics infrastructure, hence providing a model that allows a customer to migrate gradually from an on-premise analytics to a scenario with Cloud-provided analytics infrastructure. To improve the market penetration of analytics solutions in emerging markets such as India, Deepak \textit{et al.} \cite{DeepakMultiFlow:2012} propose a multi-flow solution for analytics that can be deployed on the Cloud. The multi-flow approach provides a range of possible analytics operators and flows to compose analytics solutions; viewed as workflows or instantiations of a multi-flow solution. IVOCA \cite{BhattacharyaCRM:2009} is a tool aimed at \ac{CRM} that ingests both structured and unstructured data and provides data linking, classification, and text mining tools to facilitate analysts' tasks and reduce the time to insight.

Habich~\textit{et al.}~\cite{habich2010using} propose Web services that co-ordinate data Clouds for exchanging massive data sets. The \ac{BPEL} data transition approach is used for data exchange by passing references to data between services to reduce the execution time and guarantee the correct data processing of an analytics process. A generic data Cloud layer is introduced to handle heterogeneous data Clouds, and is responsible for mapping generic operations to each Cloud implementation. DataDirect Cloud~\cite{DataDirect} also provides generic interfaces by offering JDBC/ODBC drivers for applications to execute SQL queries against different databases stored on a Cloud. Users are not required to deal with different APIs and query languages specific to each Cloud storage solution.
 
PivotLink's AnalyticsCLOUD \cite{PivotLink} handles both structured and unstructured data, providing data integration features. PivotLink also provides DataCLOUD with information about over 350 demographic, hobbies, and interest data fields for 120 million U.S. households. This information can be used by customers to perform brand sentiment analysis \cite{FeldmanSentiment:2013} and verify how weather affects their product performance. 

 

\subsection{Data Processing and Resource Management}

MapReduce \cite{JeffreyMapReduce:2008} is one of the most popular programming models to process large amounts of data on clusters of computers. Hadoop \cite{Hadoop:2012} is the most used open source MapReduce implementation, also made available by several Cloud providers \cite{AmazonEMR,barga2012project,Infochimps,HDInsight}. Amazon \ac{EMR} \cite{AmazonEMR} enables customers to instantiate Hadoop clusters to process large amounts of data using the Amazon \ac{EC2} and other Amazon Web Services for data storage and transfer.

Hadoop uses the \ac{HDFS} file system to partition and replicate data sets across multiple nodes, such that when running a MapReduce application, a mapper is likely to access data that is locally stored on the cluster node where it is executing. Although Hadoop provides a set of APIs that allow developers to implement MapReduce applications, very often a Hadoop workflow is composed of jobs that use high-level query languages such as Hive and Pig Latin, created to facilitate search and specification of processing tasks. Lee \emph{et al.}~\cite{Lee2011MapReduceSurvey} present a survey about the features, benefits, and limitations of MapReduce for parallel data analytics. They also discuss extensions proposed for this programming model to overcome some of its limitations.

Hadoop explores data parallelism, and its data and task replication enable fault tolerance, but what is often criticised about it is the time required to load data into \ac{HDFS} and the lack of reuse of data that mappers produce. MapReduce is a model created to exploit commodity hardware, but when executed on reliable infrastructure, the mechanisms it provides to deal with failures may not be entirely essential. Some of the provided features can be disabled in certain scenarios. Herodotou and Babu \cite{HerodotouWhatIf:2011} present techniques for profiling MapReduce applications, identifying bottlenecks and simulating what-if scenarios. Previous work has also proposed optimisations to handle these shortcomings \cite{GuoMapReduce:2012}. Cuzzocrea \textit{et al.} \cite{cuzzocrea2011analytics} discuss issues concerning analytics over big multidimensional data and the difficulties in building multidimensional structures in \ac{HDFS} and integrating multiple data sources to Hadoop.

Starfish \cite{herodotou2011starfish}, a data analytics system built atop Hadoop, focuses on improving the performance of clusters throughout the data lifecycle in analytics, without requiring users to understand the available configuration options. Starfish employs techniques at several levels to optimise the execution of MapReduce jobs. It uses dynamic instrumentation to profile jobs and optimises workflows by minimising the impact of data unbalance and by balancing the load of executions. Starfish's Elastisizer automates provisioning decisions using a mix of simulation and model-based estimation to address what-if questions on workload performance.

Lee \textit{et al.} \cite{LeeSchedAnalytics:2011} present an approach that allocates resources and schedules jobs considering data analytics workloads, in order to enable consolidation of a cluster workload, reducing the number of machines allocated for processing the workload during periods of small load. The approach uses Hadoop and works with two pools of machines---\textit{core} and \textit{accelerator}---and dynamically adjusts the size of each pool according to the observed load.

Daytona \cite{barga2012project}, a MapReduce runtime for Windows Azure, leverages the scalable storage services provided by Azure's Cloud infrastructure as the source and destination of data. It uses Cloud features to provide load balancing and fault tolerance. The system relies on a master-slave architecture where the master is responsible for scheduling tasks and the slaves for carrying out map and reduce operations. Section~\ref{sec:visualisation} discusses the visualisation features that Daytona provides.

Previous work shows that there is an emerging class of MapReduce applications that feature small, short, and highly interactive jobs \cite{ChenInteractive:2012,Fox:2012}. As highlighted in Section \ref{sec:visualisation}, the visualisation community often criticises the lack of interactivity of MapReduce-based analytics solutions. Over the past few years, however, several attempts have been made to tackle this issue. Borthakur \textit{et al.} \cite{BorthakurFacebookHadoop:2011}, for instance, describe optimisations implemented in \ac{HDFS} and HBase\footnote{http://hbase.apache.org} to make them more responsive to the realtime requirements of Facebook applications. Chen \textit{et al.} \cite{ChenEnergyHadoop:2012} propose energy efficiency improvements to Hadoop by maintaining two distinct pools of resources, namely to interactive and batch jobs.

The eXtreme Analytics Platform (XAP) \cite{Balmin2013XAP} enables analytics supporting multiple data sources, data types (structured and unstructured), and multiple types of analyses. The target infrastructure of the architecture is a cluster running a distributed file system. A modified version of Hadoop, deployed in the cluster, contains an application scheduler (FLEX) able to better utilise the available resources than the default Hadoop scheduler. The analytics jobs are created via a high-level language script, called Jaql, that converts the high-level descriptive input into an analytics MapReduce workflow that is executed in the target infrastructure.

Previous work has also considered other models for performing analytics, such as scientific workflows and \ac{OLAP}. Rahman \textit{et al.} \cite{RahmanWorkflow:2011} propose a hybrid heuristic for scheduling data analytics workflows on heterogeneous Cloud environments; a heuristic that optimises cost of workflow execution and satisfies users requirements, such as budget, deadline, and data placement.

In the field of simulation-enabled analytics, Li \emph{et al.}~\cite{Li2012Workflow}  developed an analytical application, modelled as a \ac{DAG}, for predicting the spread of dengue fever outbreaks in Singapore. The analytics workflow receives data from multiple sources, including current and past data about climate and weather from meteorological agencies and historical information about dengue outbreaks in the country. This data, with user-supplied input about the origin of the infection, is used to generate a map of the spread of the disease in the country in a day-by-day basis. A hybrid Cloud is used to speed up the application execution. Other characteristics of the application are security features and cost-effective exploration of Cloud resources: the system keeps the utilisation of public Cloud resources to a minimum to enable the analytics to complete in the specified time and budget. A public Cloud has also been used in a similar scenario to simulate the impact of public transport disruptions on urban mobility \cite{Scale:2013}.


Chohan \textit{et al.} \cite{ChohanHybridCloud:2012} evaluated the support of \ac{OLAP} for \ac{GAE} \cite{GoogleAppEngine:2012} highlighting limitations and assessing their impact on cost and performance of applications. A hybrid approach to perform \ac{OLAP} using \ac{GAE} and AppScale \cite{BunchAppScale:2010} was provided, using two methods for data synchronisation, namely \textit{bulk data transfer} and \textit{incremental data transfer}. Moreover, Jung~\textit{et al.}~\cite{JungOverlap:2012} propose optimisations for scheduling and processing of Big Data analysis on federated Clouds. 

Chang~\textit{et al.}~\cite{chang2012workload} examined different data analytics workloads, where results show significant diversity of resource usage (CPU, I/O and, network). They recommend the use of transformation mechanisms such as indexing, compression, and approximation to provide a balanced system and improve efficiency of data analysis.

The Cloud can also be used to extend the capabilities of analyses initially started on the customer's premises. CloudComet, for example, is an autonomic computing engine that supports Cloud bursts that has been used to provide the programming and runtime infrastructure to scale out/in certain on-line risk analyses \cite{KimRisk:2009}. CloudComet and commercial technologies such as Aneka~\cite{aneka} can utilise both private resources and resources from a public Cloud provider to handle peaks in the demands of online risk analytics.

Some analytics applications including stock quotes and weather prediction have stringent time constraints, usually falling in the near-time and stream categories described earlier. Request processing time is important to deliver results in a timely fashion. Chen \textit{et al.} \cite{ChenCaaaS:2011} investigate \ac{CaaaS} that blends stream processing and relational data techniques to extend the DBMS model and enable real-time continuous analytics service provisioning. The dynamic stream processing and static data management for data intensive analytics are unified by providing an SQL-like interface to access both static and stream data. The proposed cycle-based query model and transaction model allow SQL queries to run and to commit per cycle whilst analysing stream data per chunk. The analysis results are made visible to clients whilst a continued query for results generation is still running. Existing work on stream and near-time processing attempt to leverage strategies to predict user or service behaviour \cite{yunwen2011methods}. In this way, an analytics service can pre-fetch data to anticipate a user's behaviour, hence selecting the appropriate applications and methods before the user's request arrives. 

Realtime analysis of Big Data is a hot topic, with Cloud providers increasingly offering solutions that can be used as building blocks of stream and complex event processing systems. AWS Kinesis \cite{AmazonKinesis} is an elastic system for real-time processing of streaming data that can handle multiple sources, be used to build dashboards, handle events, and generate alerts. It allows for integration with other AWS services. In addition, stream processing frameworks including Apache S4 \cite{ApacheS4}, Storm \cite{ApacheStorm} and IBM InfoSphere Streams \cite{IBMStreams} can be deployed on existing Cloud offerings. Software systems such as \textit{storm-deploy}, a Clojure project based on Pallet\footnote{http://github.com/pallet/pallet}, aim to ease deployment of Storm topologies on Cloud offerings including AWS EC2. Suro, a data pipeline system used by Netflix to collect events generated by its applications, has recently been made available to the broader community as an open source project \cite{NetflixSuro}. Aiming to address similar requirements, Apache Kafka \cite{GoodhopeKafka2011} is a real-time publish-subscribe infrastructure initially used at LinkedIn to process activity data and later released as an open source project. Incubated by the Apache Software Foundation, Samza \cite{ApacheSamza} is a distributed stream processing framework that blends Kafka and Apache Hadoop YARN. Whilst Samza provides a model where streams are the input and output to jobs, execution is completely handled by YARN.

%

%


\subsection{Challenges in Big Data Management}

In this section, we discussed current research targeting the issue of Big Data management for analytics. There are still, however, many open challenges in this topic. The list below is not exhaustive, and as more research in this field is conducted, more challenging issues will arise.

\begin{description}
\item[Data variety, volume and velocity:] How to handle an always increasing volume of data? Especially when the data is unstructured, how to quickly extract meaningful content out of it? How to aggregate and correlate streaming data from multiple sources?

\item[Data storage:] How to efficiently recognise and store important information extracted from unstructured data? How to store large volumes of information in a way it can be timely retrieved? Are current file systems optimised for the volume and variety demanded by analytics applications? If not, what new capabilities are needed? How to store information in a way that it can be easily migrated/ported between data centers/Cloud providers?

\item[Data integration:] New protocols and interfaces for integration of data that are able to manage data of different nature (structured, unstructured, semi-structured) and sources.

\item[Data Processing and Resource Management:] New programming models optimised for streaming and/or multidimensional data; new backend engines that manage optimised file systems; engines able to combine applications from multiple programming models (\textit{e.g.} MapReduce, workflows, and bag-of-tasks) on a single solution/abstraction. How to optimise resource usage and energy consumption when executing the analytics application?
\end{description}

\section{Model Building and Scoring}
\label{sec:models}

The data storage and \ac{DaaS} capabilities provided by Clouds are important, but for analytics, it is equally relevant to use the data to build models that can be utilised for forecasts and prescriptions. Moreover, as models are built based on the available data, they need to be tested against new data in order to evaluate their ability to forecast future behaviour. Existing work has discussed means to offload such activities---termed here as model building and scoring---to Cloud providers and ways to parallelise certain machine learning algorithms \cite{UpadhyayaMachineLearning:2013,ApacheMahout,HuangCumulon:2013}. This section describes work on the topic. Table \ref{tab:modelBuilding} summarises the analysed work, its goals, and target infrastructures.

\begin{table}
\centering
\caption{Summary of works on model building and scoring.}
\label{tab:modelBuilding}
\begin{footnotesize}
\begin{tabular}{p{43mm}lcc}
\toprule
\multirow{2}{43mm}{\centering{\bf Work}} & 
\multirow{2}{30mm}{\centering{\bf Goal}}  & 
\multirow{2}{12mm}{\centering{\bf Service model}}  & 
\multirow{2}{20mm}{\centering{\bf Deployment Model}} \\ 
& & & \\
\toprule
Guazzelli \textit{et al.} \cite{GuazzelliADAPA:2009}	&  Predictive analytics (scoring) 	 & IaaS	 & Public \\ 
\midrule
\multirow{2}{43mm}{Zementis \cite{Zementis:2012}}	 &  \multirow{2}{30mm}{Data Analysis and Model Building} & \multirow{2}{12mm}{\centering{SaaS}}	& \multirow{2}{18mm}{\centering{Public or Private}} \\ 
 & & & \\
 \midrule
Google Prediction API \cite{GooglePredictionAPI}	& Model Building				& SaaS	& Public \\ 
\midrule
\multirow{2}{43mm}{Apache Mahout \cite{ApacheMahout}}	 &  \multirow{2}{30mm}{Data Analysis and Model Building} & \multirow{2}{12mm}{\centering{IaaS}}	& \multirow{2}{18mm}{\centering{Any}} \\ 
 & & & \\
\midrule
Hazy \cite{KumarHazy:2013}					& Model Building			 	& IaaS	& Any \\ 
\bottomrule
\end{tabular}
\end{footnotesize}
\end{table}

Guazzelli \textit{et al.} \cite{GuazzelliADAPA:2009} use Amazon \ac{EC2} as a hosting platform for the Zementis' ADAPA model \cite{Zementis:2012} scoring engine. Predictive models, expressed in \ac{PMML} \cite{GuazzelliPMML:2009}, are deployed in the Cloud and exposed via Web Services interfaces. Users can access the models with Web browser technologies to compose their data mining solutions. Existing work also advocates the use of \ac{PMML} as a language to exchange information about predictive models \cite{HsuehWelness:2010}.

Zementis \cite{Zementis:2012} also provides technologies for data analysis and model building that can run either on a customer's premises or be allocated as \ac{SaaS} using \ac{IaaS} provided by solutions such as Amazon \ac{EC2} and IBM SmartCloud Enterprise \cite{SmartCloud:2012}.

Google Prediction API \cite{GooglePredictionAPI} allows users to create machine learning models to predict numeric values for a new item based on values of previously submitted training data or predict a category that best describes an item. The prediction API allows users to submit training data as comma separated files following certain conventions, create models, share their models or use models that others shared. With the Google Prediction API, users can develop applications to perform analytics tasks such as sentiment analysis \cite{FeldmanSentiment:2013}, purchase prediction, provide recommendations, analyse churn, and detect spam. The Apache Mahout project \cite{ApacheMahout} aims to provide tools to build scalable machine learning libraries on top of Hadoop using the MapReduce paradigm. The provided libraries can be deployed on a Cloud and be explored to build solutions that require clustering, recommendation mining, document categorisation, among others.

By trying to ease the complexity in building trained systems such as IBM's Watson, Apple's Siri and Google Knowledge Graph, the Hazy project \cite{KumarHazy:2013} focuses on identifying and validating two categories of abstractions in building trained systems, namely \textit{programming abstractions} and \textit{infrastructure abstractions}. It is argued that, by providing such abstractions, it would be easier for one to assemble existing solutions and build trained systems. To achieve a small and compoundable programming interface, Hazy employs a data model that combines the relational data model and a probabilistic rule-based language. For infrastructure abstraction, Hazy leverages the observation that many statistical analysis algorithms behave as a user-defined aggregate in a \ac{RDBMS}. Hazy then explores features of the underlying infrastructure to improve the performance on these aggregates.

\subsection{Open Challenges}

The key challenge in the area of Model Building and Scoring is the discovery of techniques that are able to explore the rapid elasticity and large scale of Cloud systems. Given that the amount of data available for Big Data analytics is increasing, timely processing of such data  for building and scoring would give a relevant advantage for businesses able to explore such a capability.

In the same direction, standards and interfaces for these activities are also required, as they would help to disseminate ``prediction and analytics as services'' providers that would compete for customers. If the use of such services does not incur vendor lock in (via utilisation of standards APIs and formats), customers can choose the service provider only based on cost and performance of services, enabling the emergence of a new competitive market.

\section{Visualisation and User Interaction}
\label{sec:visualisation}

With the increasing amounts of data with which analyses need to cope, good visualisation tools are crucial. These tools should consider the quality of data and presentation to facilitate navigation \cite{davey2012visual}. The type of visualisation may need to be selected according to the amount of data to be displayed, to improve both displaying and performance. Visualisation can assist in the three major types of analytics: descriptive, predictive, and prescriptive. Many visualisation tools do not describe advanced aspects of analytics, but there has been an effort to explore visualisation to help on predictive and prescriptive analytics, using for instance sophisticated reports and storytelling \cite{kosara2013storytelling}. A key aspect to be considered on visualisation and user interaction in the Cloud is that network is still a bottleneck in several scenarios \cite{tabor2013ubercloud}. Users ideally would like to visualise data processed in the Cloud having the same experience and feel as though data were processed locally. Some solutions have been tackling this requirement.

For example, as Fisher \textit{et al.} \cite{fisher2012interactions} point out, many Cloud platforms available to process data analytics tasks still resemble the \textit{batch-job} model used in the early times of the computing era. Users typically submit their jobs and wait until the execution is complete to download and analyse sample results to validate full runs. As this back and forth of data is not well supported by the Cloud, the authors issue a call to arms for both research and development of better interactive interfaces for Big Data analytics where users iteratively pose queries and see rapid responses. Fisher \textit{et al.} introduce \textit{sampleAction} \cite{FisherTrustMet:2012} to explore whether interactive techniques acting over only incremental samples can be considered as sufficiently trustworthy by analysts to make closer to real time decisions about their queries. Interviews with three teams of analysts suggest that representations of incremental query results were robust enough so that analysts were prepared either to abandon a query, refine it, or formulate new queries. King \cite{KingBuyDataMining:2008} also highlights the importance of making the analytics process iterative, with multiple checkpoints for assessment and adjustment.

In this line, existing work aims to explore the batch-job model provided by solutions including MapReduce as a backend to features provided in interfaces with which users are more familiar. Trying to leverage the popularity of spreadsheets as a tool to manipulate data and perform analysis, Barga \textit{et al.} proposed an Excel ribbon connected to Daytona \cite{barga2012project}, a Cloud service for data storage and analytics. Users manipulate data sets on Excel and plugins use Microsoft's Azure infrastructure \cite{CalderAzure:2011} to run MapReduce applications. In addition, as described earlier, several improvements have been proposed to MapReduce frameworks to handle interactive applications  \cite{BorthakurFacebookHadoop:2011,ChenEnergyHadoop:2012,MelnikDremel:2010}. However, most of these solutions are not yet made available for general use in the Cloud.

Several projects attempt to provide a range of visualisation methods from which users can select a set that suits their requirements. ManyEyes~\cite{viegas2007manyeyes} from IBM allows users to upload their data, select a visualisation method---varying from basic to advanced---and publish their results. Users may also navigate through existing visualisations and discuss their findings and experience with peers. Selecting data sources automatically or semi-automatically is also an important feature to help users perform analytics. PanXpan~\cite{panxpan} is an example of a tool that automatically identifies the fields in structured data sets based on user analytics module selection. FusionCharts~\cite{fusioncharts} is another tool to allow users to visually select a subset of data from the plotted data points to be submitted back to the server for further processing. CloudVista~\cite{chen2011cloudvista,xu2012cloudvista} is a software to help on visual data selection for further analysis  refinement.

Existing work also provides means for users to aggregate data from multiple sources and employ various visualisation models, including dashboards, widgets, line and bar charts, demographics, among other models \cite{PivotLink,lu2011framework,GoogleAnalytics,Gooddata,VisualizationWeb:2013}. Some of these features can be leveraged to perform several tasks, including create reports; track what sections of a site are performing well and what kind of content can create better user experience; how information sharing on a social network impacts the web site usage; track mobile usage \cite{attentionshoppers:2013,CiscoMSE:2012}; and evaluate the impact of advertising campaigns.

Choo and Park \cite{choo:2013} argue that the reason why Big Data visualisation is not real time is the computational complexity of the analytics operations. In this direction, authors discuss strategies to reduce computational complexity of data analytics operations by, for instance, decreasing precision of calculations.

Apart from software optimisation, dedicated hardware for visualisation is becoming key for Big Data analytics. For example, Reda \emph{et al.} \cite{Reda2013CAVE2} discuss that, although existing tools are able to provide data belonging to a range of classes, their dimensionality and volume exceed the capacity of visualisation provided by standard displays. This requires the utilisation of large-scale visualisation environments, such as CyberCommons and CAVE2, which are composed of a large display wall with resolution three orders of magnitude higher than that achieved by commercial displays~ \cite{Reda2013CAVE2}. Remote visualisation systems, such as Nautilus from XSEDE (Extreme Science and Engineering Discovery Environment---the new NSF TeraGrid project replacement), are becoming more common to supply high demand for memory and graphical processors to assist very large data visualisation \cite{xsede}.

Besides visualisation of raw data, summarised content in form of reports are essential to perform predictive and prescriptive analytics. Several solutions have explored report generation and visualisation. For instance, SAP Crystal Solutions \cite{SAPCrystal} provides \ac{BI} functionalities via which customers can explore available data to build reports with interactive charts, what-if scenarios, and dashboards. The produced reports can be visualised on the Web, e-mail, Microsoft Office, or be embedded into enterprise applications. Another example on report visualisation is Cloud9 Analytics \cite{Cloud9}, which aims to automate reports and dashboards, based on data from \ac{CRM} and other systems. It provides features for sales reports, sales analytics, and sales forecasts and pipeline management. By exploring history data and using the notion of risk, it offers customers clues on which projects they should invest their resources and what projects or products require immediate action. Other companies also offer solutions that provide sales forecasts, change analytics, and customised reports \cite{Right90,Birst}. Salesforce \cite{SalesForce} supports customisable dashboards through collaborative analytics. The platform allows authorised users to share their charts and information with other users. Another trend on visualisation to help on predictive and prescriptive analytics is storytelling \cite{kosara2013storytelling}, which aims at presenting data with a narrative visualisation.

There are also visualisation tools specific for a given domain. For instance, in the field of climate-modelling, Lee \emph{et al.}~\cite{Lee2013MJO} developed a tool for visualisation of simulated Madden-Julian Oscillation, which is an important meteorological event that influences raining patterns from South America to Southeast Asia. The tool enables tracking of the event and its visualisation using Google Earth. In the area of computing networks management, Liao \emph{et al.}~\cite{liao:13} evaluated five approaches for visualisation of anomalies in large scale computer networks. Each method has its own applications depending on the specific type of anomaly to be visualised and the scale of the managed system. There are also solutions that provide means to visualise demographic information. Andrienko~\textit{et al.}~\cite{andrienko2007visual} proposed interactive visual display for analysis of movement behaviour of people, vehicle, and animals. The visualisation tool displays the movement data, information about the time spent in a place, and the time interval from one place to another.

%
%
%
%

\subsection{Open Challenges}

There are many research challenges in the field of Big Data visualisation. First, more efficient data processing techniques are required in order to enable real-time visualisation.  Choo and Park \cite{choo:2013} appoint some techniques that can be employed with this objective, such as reduction of accuracy of results, coarsely processing of data points, compatible with the resolution of the visualisation device, reduced convergence, and data scale confinement. Methods considering each of these techniques could be further researched and improved.

Cost-effective devices for large-scale visualisation is another hot topic for analytics visualisation, as they enable finer resolution than simple screens. Visualisation for management of computer networks and software analytics~\cite{ieeeComputer201303} are also areas that are attracting attention of researchers and practitioners for its extreme relevance to management of large-scale infrastructure (such as Clouds) and  software, with implications in global software development, open source software development, and software quality improvements.

\section{Business Models and Non-Technical Challenges}
\label{sec:business}

In addition to providing tools that customers can use to build their Big Data analytics solutions on the Cloud, models for delivering analytics capabilities as services on a Cloud have been discussed in previous work \cite{SunBusinessModels:2011}. Sun \textit{et al.}~\cite{SunBusinessModels:2012} provide an overview of the current state of the art on the development of customised analytics solutions on customers' premises and elaborate on some of the challenges to enable analytics and analytics as a service on the Cloud. Some of the potential business models proposed in their work include:

\begin{itemize}
\item \textbf{Hosting customer analytics jobs in a shared platform:} suitable for an enterprise or organisation that has multiple analytics departments. Traditionally, these departments have to develop their own analytics solutions and maintain their own clusters. With a shared platform they can upload their solutions to execute on a shared infrastructure, therefore reducing operation and maintenance costs. As discussed beforehand, techniques have been proposed for resource allocation and scheduling of Big Data analytics tasks on the Cloud \cite{LeeSchedAnalytics:2011,RahmanWorkflow:2011}.

\item \textbf{A full stack designed to provide customers with end-to-end solutions:} appropriate for companies that do not have expertise on analysis. In this model, analytical service providers publish domain-specific analytical stream templates as services. The provider is responsible for hosting the software stack and managing the resources necessary to perform the analyses. Customers who subscribe to the services just need to upload their data, configure the templates, receive models, and perform the proper model scoring.

\item \textbf{Expose analytics models as hosted services:} analytics capabilities are hosted on the Cloud and exposed to customers as services. This model is proposed to companies that do not have enough data to make good predictions. Providers upload their models, which are consumed by customers via scoring services provided by the Cloud.
\end{itemize}

To make Big Data analytics solutions more affordable, Sun \textit{et al.} \cite{SunBusinessModels:2012} also propose cost-effective approaches that enable multi-tenancy at several levels. They discuss the technical challenges on isolating analytical artefacts. Hsueh \textit{et al.} \cite{HsuehWelness:2010} discuss issues related to pricing and \acp{SLA} on a platform for personalisation in a wellness management platform built atop a Cloud infrastructure. Krishna and Varma \cite{Krishnainfosys:2012} envision two types of services for Cloud analytics: (i) \ac{AaaS}, where analytics is provided to clients on demand and they can pick the solutions required for their purposes; and (ii) \ac{MaaS} where models are offered as building blocks for analytics solutions.

Bhattacharya \textit{et al.} \cite{BhattacharyaCRM:2009} introduced IVOCA, a solution for providing managed analytics services for \ac{CRM}. IVOCA provides functionalities that help analysts better explore data analysis tools to reduce the time to insight and improve the repeatability of \ac{CRM} analytics. Also in the \ac{CRM} realm, KXEN \cite{KXEN} offers a range or products for performing analytics, some of which can run on the Cloud. Cloud Prediction is a predictive analytics solution for Salesforce.com. With its Predictive Lead Scoring, Predictive Offers, and Churn Prediction, customers can leverage the \ac{CRM}, mobile, and social data available in the Cloud to score leads based on which ones can create sales opportunities; create offers that have a higher likelihood to be accepted based on a prediction of offers and promotions; and gain insights into which customers a company is at risk of losing.

Cloud-enabled Big Data analytics poses several challenges in respect to replicability of analyses. When not delivered by a Cloud, analytics solutions are customer-specific and models often have to be updated to consider new data. Cloud solutions for analytics need to balance generality and usefulness. Previous work also discusses the difficulty of replicating activities of text analytics \cite{ProctorBIConsulting:2011}. An analytical pathway is proposed to link business objectives to an analytical flow, with the goal of establishing a methodology that illustrates and possibly supports repeatability of analytical processes when using complex analytics. King \cite{KingBuyDataMining:2008}, whilst discussing some of the problems in buying predictive analytics, provides a best practice framework based on five steps, namely training, assessment, strategy, implementation, and iteration.

Chen \textit{et al.} \cite{ChenEcosystem:2011} envision an analytics ecosystem where data services aggregate, integrate, and provide access to public and private data by enabling partnerships among data providers, integrators, aggregators, and clients; these services are termed as \ac{DaaS}. Atop \ac{DaaS}, a range of analytics functionalities that explore the data services are offered to customers to boost productivity and create value. This layer is viewed as \ac{AaaS}. Similar to the previously described work, they discuss a set of possible business models that range from proprietary, where both data and models are kept private, to co-developing models where both data and analytics models are shared among the parties involved in the development of the analytics strategy or services.




\section{Other Challenges}
\label{sec:gap_analysis}

In business models where high-level analytics services may be delivered by the Cloud, human expertise cannot be easily replaced by machine learning and Big Data analysis \cite{McAfeeBigData:2012}; in certain scenarios, there may be a need for human analysts to remain in the loop \cite{LazerGoogleFlu2014}. Management should adapt to Big Data scenarios and deal with challenges such as how to assist human analysts in gaining insights and how to explore methods that can help managers in making quicker decisions.

Application profiling is often necessary to estimate the costs of running analytics on a Cloud platform. Users need to develop their applications to target Cloud platforms; an effort that should be carried out only after estimating the costs of transferring data to the Cloud, allocating virtual machines, and running the analysis. This cost estimation is not a trivial task to perform in current Cloud offerings. Although best practices for using some data processing services are available \cite{BestPracticesEMR:2013}, there should be tools that assist customers to estimate the costs and risks of performing analytics on the Cloud.

Data ingestion by Cloud solutions is often a weak point, whereas debugging and validating developed solutions is a challenging and tedious process. As discussed earlier, the manner analytics is executed on Cloud platforms resembles the batch job scenario: users submit a job and wait until tasks are executed to then download the results. Once an analysis is complete, they download sample results that are enough to validate the analysis task and after that perform further analysis. Current Cloud environments lack this interactive process, and techniques should be developed to facilitate interactivity and to include analysts in the loop by providing means to reduce their time to insight. Systems and techniques that iteratively refine answers to queries and give users more control of processing are desired \cite{HellersteinCONTROL:1999}.

Furthermore, market research shows that inadequate staffing and skills, lack of business support, and problems with analytics software are some of the barriers faced by corporations when performing analytics \cite{RussomBigDataAnalytics:2011}. These issues can be exacerbated by the Cloud as the resources and analysts involved in certain analytics tasks may be offered by a Cloud provider and may move from one customer engagement to another. In addition, based on survey responses, currently most analytics updates and scores of methods occur daily to annually; which can become an issue for analytics on streaming data. Russom \cite{RussomBigDataAnalytics:2011} also highlights the importance of advanced data visualisation techniques and advanced analytics---such as analysis of unstructured, large data sets and streams---to organisations in the next few years.

Chen \textit{et al.} \cite{ChenBI30:2012} foresee the emergence of what they termed as Business Intelligence and Analytics (BI\&A) 3.0, which will require underlying mobile analytics and location and context-aware techniques for collecting, processing, analysing, and visualising large scale mobile and sensor data. Many of these tools are still to be developed. Moreover, moving to BI\&A 3.0 will demand efforts on integrating data from multiple sources to be processed by Cloud resources, and using the Cloud to assist decisions by mobile device users. 

More recently, terms such as Analytics as a Service (AaaS) and Big Data as a Service (BDaaS) are becoming popular. They comprise services for data analysis similarly as IaaS offers computing resources. However, these analytics services still lack well defined contracts since it may be difficult to measure quality and reliability of results and input data, provide promises on execution times, and guarantees on methods and experts responsible for analysing the data. Therefore, there are fundamental gaps on tools to assist service providers and clients to perform these tasks and facilitate the definition of contracts for both parties.

\section{Summary and Conclusions}
\label{sec:conclusions}

The amount of data currently generated by the various activities of the society has never been so big, and is being generated in an ever increasing speed. This Big Data trend is being seen by industries as a way of obtaining advantage over their competitors: if one business is able to make sense of the information contained in the data reasonably quicker, it will be able to get more costumers, increase the revenue per customer, optimise its operation, and reduce its costs. Nevertheless, Big Data analytics is still a challenging and time demanding task that requires expensive software, large computational infrastructure, and effort.

Cloud computing helps in alleviating these problems by providing resources on-demand with costs proportional to the actual usage. Furthermore, it enables infrastructures to be scaled up and down rapidly, adapting the system to the actual demand.

Although Cloud infrastructure offers such elastic capacity to supply computational resources on demand, the area of Cloud-supported analytics is still in its early days. In this paper, we discussed the key stages of analytics workflows, and surveyed the state-of-the-art of each stage in the context of Cloud-supported analytics. Surveyed work was classified in three key groups: Data Management  (which encompasses data variety, data storage, data integration solutions, and data processing and resource management), Model Building and Scoring, and Visualisation and User Interactions. For each of these areas, ongoing work was analysed and key open challenges were discussed. This survey concluded with an analysis of business models for Cloud-assisted data analytics and other non-technical challenges.

The area of Big Data Computing using Cloud resources is moving fast, and after surveying the current solutions we identified some key lessons:
\begin{itemize}

\item There are plenty of solutions for Big Data related to Cloud computing. Such a large number of solutions has been created because of the wide range of analytics requirements, but they may, sometimes, overwhelm non-experienced users. Analytics can be descriptive, predictive, prescriptive; Big Data can have various levels of variety, velocity, volume, and veracity. Therefore, it is important to understand the requirements in order to choose appropriate Big Data tools;

\item It is also clear that analytics is a complex process that demands people with expertise in cleaning up data, understanding and selecting proper methods, and analysing results. Tools are fundamental to help people perform these tasks. In addition, depending on the complexity and costs involved in carrying out these tasks, providers who offer Analytics as a Service or Big Data as a Service can be a promising alternative compared to performing these tasks in-house;

\item Cloud computing plays a key role for Big Data; not only because it provides infrastructure and tools, but also because it is a business model that Big Data analytics can follow (\textit{e.g.} Analytics as a Service (AaaS) or Big Data as a Service (BDaaS)). However, AaaS/BDaaS brings several challenges because the customer and provider's staff are much more involved in the loop than in traditional Cloud providers offering infrastructure/platform/software as a service.

\end{itemize}

Recurrent themes among the observed future work include (i) the development of standards and APIs enabling users to easily switch among solutions and (ii) the ability of getting the most of the elasticity capacity of the Cloud infrastructure. The latter includes expressive languages that enable users to describe the problem in simple terms whilst decomposing such high-level description in highly concurrent subtasks and keeping good performance efficiency even for large numbers of computing resources. If this can be achieved, the only limitations for an arbitrary short processing time would be market issues, namely the relation between the cost for running the analytics and the financial return brought for the obtained knowledge.

\bibliographystyle{elsarticle-num}
\bibliography{references}

\end{document}